\documentclass[sigconf]{acmart}

\usepackage{booktabs} 
\usepackage{hyperref}
\usepackage{url}

\usepackage{times}
\usepackage{amssymb}
\usepackage{dirtytalk}

\usepackage{tablefootnote}

\usepackage{enumitem}
\usepackage{xcolor}
\usepackage{colortbl}
\usepackage{graphicx}
\usepackage{subcaption}
\usepackage[font=small]{caption}
\usepackage{multirow}
\usepackage{microtype}
\usepackage{multicol}
\usepackage{tabularx} 
\usepackage{graphics}
\usepackage{adjustbox}
\usepackage{csquotes}
\usepackage{nicefrac}
\usepackage{booktabs}
\usepackage{setspace}
\usepackage{pgfplotstable} 
\usepackage{graphicx}
\usepackage{pgfplots}
\usepgfplotslibrary{statistics}
\usepackage{graphics}
\usepackage{adjustbox}

\usetikzlibrary{pgfplots.groupplots}
\pgfplotsset{compat=1.11}
\usepgfplotslibrary{ternary}
\usepackage{tikz}

\usepackage{algorithm}
\usepackage{algorithmic}

\usetikzlibrary{arrows.meta}
\usepackage{tkz-graph}
\usetikzlibrary{positioning,automata}
\usetikzlibrary{calc,spy,shapes}
\usetikzlibrary{arrows,decorations.pathmorphing,fit,positioning,patterns}
\pgfplotsset{compat=newest}
\DeclareMathOperator*{\argmin}{arg\,min}

\newcommand{\mypar}[1]{\noindent\textbf{#1}~}

\newcommand{\jk}[1]{\textcolor{blue}{#1}}
\renewcommand{\jk}[1]{}

\usepackage{xspace} 

\newcommand{\fwl}{FWL\xspace}

\newcommand{\fwlfull}{Fidelity-Weighted Learning\xspace}
\newcommand{\fwlfulllc}{fidelity-weighted learning\xspace}

\newcommand{\std}{student\xspace}
\newcommand{\tch}{teacher\xspace}

\newcommand{\wa}{weak annotator\xspace}
\newcommand{\was}{weak annotators\xspace}

\newcommand{\tfunc}{T} 

\DeclareFontFamily{U}{mathb}{\hyphenchar\font45}
\DeclareFontShape{U}{mathb}{m}{n}{
<-6> mathb5 <6-7> mathb6 <7-8> mathb7
<8-9> mathb8 <9-10> mathb9
<10-12> mathb10 <12-> mathb12
}{}
\DeclareSymbolFont{mathb}{U}{mathb}{m}{n}

\DeclareMathSymbol{\smalltriangleup} {2}{mathb}{"98}
\DeclareMathSymbol{\smalltriangledown} {2}{mathb}{"99}
\DeclareMathSymbol{\smalltriangleleft} {2}{mathb}{"9A}
\DeclareMathSymbol{\smalltriangleright}{2}{mathb}{"9B}
\DeclareMathSymbol{\blacktriangleup} {2}{mathb}{"9C}
\DeclareMathSymbol{\blacktriangledown} {2}{mathb}{"9D}
\DeclareMathSymbol{\blacktriangleleft} {2}{mathb}{"9E}
\DeclareMathSymbol{\blacktriangleright}{2}{mathb}{"9F}
\def\:{\hskip0pt} 
\usepackage{siunitx}  
\setcopyright{acmlicensed}
\acmConference[LND4IR '18]{ACM SIGIR 2018 Workshop on Learning from Limited or Noisy Data}{July 12, 2018}{Ann Arbor, Michigan, USA} 
\acmYear{2018}
\copyrightyear{2018}
\acmDOI{}
\acmISBN{}
\acmPrice{}

\author{Mostafa Dehghani}
\affiliation{%
  \institution{University of Amsterdam}
}
\email{dehghani@uva.nl}

\author{Jaap Kamps}
\affiliation{%
  \institution{University of Amsterdam}
}
\email{kamps@uva.nl}

\begin{document}

\title{Learning to Rank from Samples of Variable Quality}

\begin{abstract}
Training deep neural networks requires many training samples, but in practice training labels are expensive to obtain and may be of varying quality, as some may be from trusted expert labelers while others might be from heuristics or other sources of weak supervision such as crowd-sourcing.
This creates a fundamental quality-versus-quantity trade-off in the learning process. Do we learn from the small amount of high-quality data or the potentially large amount of weakly-labeled data?
We argue that if the learner could somehow know and take the label-quality into account when learning the data representation, we could get the best of both worlds. 
To this end, we introduce ``\fwlfulllc'' (\fwl)~\citep{Dehghani:2018fidelity}, a semi-supervised student-teacher approach for training deep neural networks using weakly-labeled data. \fwl modulates the parameter updates to a \emph{student} network (trained on the task we care about) on a per-sample basis according to the posterior confidence of its label-quality estimated by a \emph{teacher} (who has access to the high-quality labels).  Both student and teacher are learned from the data. We evaluate \fwl on document ranking where we outperform state-of-the-art alternative semi-supervised methods.
\end{abstract}

\maketitle

\section{Introduction}
\label{sec:introduction}
The success of deep neural networks to date depends strongly on the availability of labeled data which is costly and not always easy to obtain. 
Usually it is much easier to obtain small quantities of high-quality labeled data and large quantities of unlabeled data. The problem of how to best integrate these two different sources of information during training is an active pursuit in the field of semi-supervised learning~\citep{chap:semi06}.
However, for a large class of tasks it is also easy to define one or more so-called ``weak annotators'', additional (albeit noisy) sources of \emph{weak supervision} based on heuristics or ``weaker'', biased classifiers trained on e.g.\ non-expert crowd-sourced data or data from different domains that are related. While easy and cheap to generate, it is not immediately clear if and how these additional weakly-labeled data can be used to train a stronger classifier for the task we care about.
More generally, in almost all practical applications machine learning systems have to deal with data samples of variable quality. For example, in a large dataset of images only a small fraction of samples may be labeled by experts and the rest may be crowd-sourced using e.g.\ Amazon Mechanical Turk.

Assuming we can obtain a large set of weakly-labeled data in addition to a much smaller training set of ``strong'' labels, 
the simplest approach is to expand the training set by including the weakly-supervised samples (all samples are equal). Alternatively, one may pretrain on the weak data and then fine-tune on observations from the true function or distribution (which we call strong data). Indeed, it has recently been shown that a small amount of expert-labeled data can be augmented in such a way by a large set of raw data, with labels coming from a heuristic function, to train a more accurate neural ranking model~\citep{Dehghani:2017:SIGIR, Dehghani:2017avoiding, Dehghani:2017:nips_metalearn}.
The downside is that such approaches are oblivious to the amount or source of noise in the labels. Simply speaking, they do not consider the cause of noise in the labels and only focus on the effect. 

In this paper, we argue that treating weakly-labeled samples uniformly (i.e.\ each weak sample contributes equally to the final classifier) ignores potentially valuable information of the label quality. Instead, we propose \fwlfull (\fwl), a Bayesian semi-supervised approach that leverages a small amount of data with true labels to generate a larger training set with \emph{confidence-weighted weakly-labeled samples}, which can then be used to modulate the fine-tuning process based on the fidelity (or quality) of each weak sample. By directly modeling the inaccuracies introduced by the \wa in this way, we can control the extent to which we make use of this additional source of weak supervision: more for confidently-labeled weak samples close to the true observed data, and less for uncertain samples further away from the observed data. 

\begin{figure*}[!t]%
    \makebox[\textwidth][c]{
    \centering
    \begin{subfigure}[t]{0.325\textwidth}
        \centering
        \includegraphics[width=\textwidth]{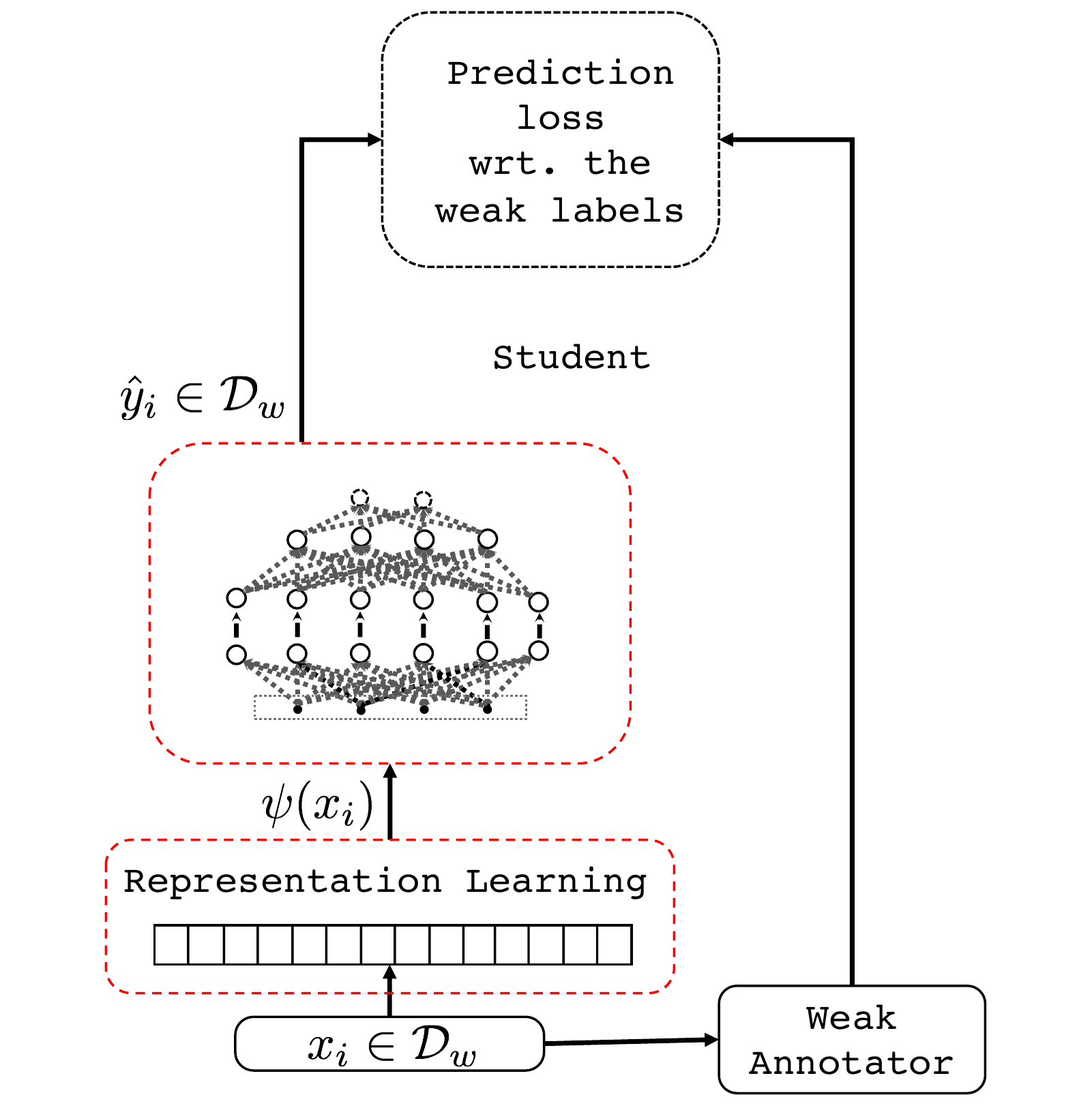}
        \caption{\label{fig:step1}\footnotesize{Step 1}}
        \vspace{-5pt}
    \end{subfigure}%
    ~
    \begin{subfigure}[t]{0.25\textwidth}
        \centering
        \includegraphics[width=\textwidth]{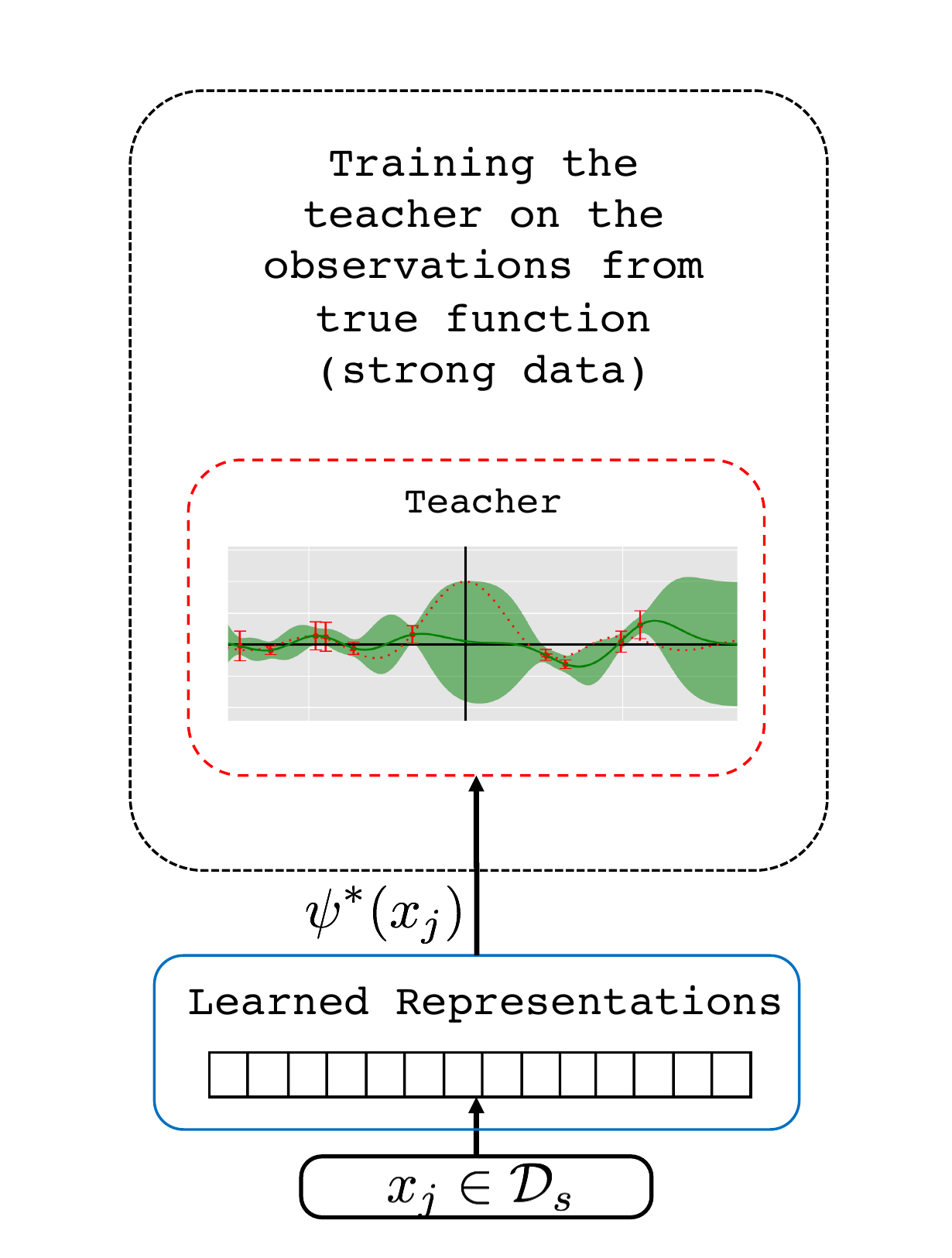}
        \caption{\label{fig:step2}\footnotesize{Step 2}}
        \vspace{-5pt}
    \end{subfigure}%
     ~
    \begin{subfigure}[t]{0.425\textwidth}
        \centering
        \includegraphics[width=\textwidth]{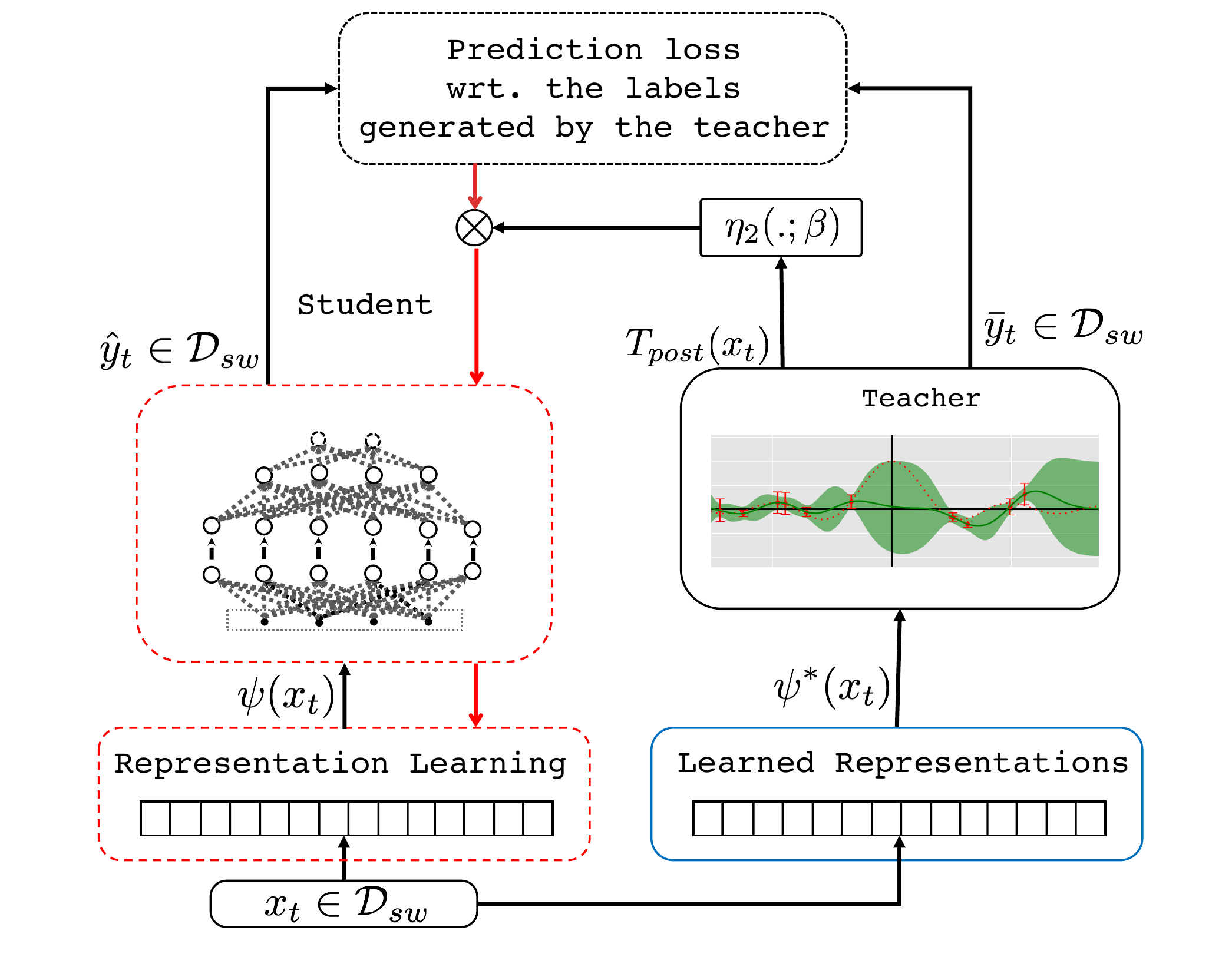}
        \caption{\label{fig:step3}\footnotesize{Step 3}}
        \vspace{-5pt}
    \end{subfigure}%
    }
    \caption{\fontsize{8}{7}\selectfont{Illustration of \fwlfull: Step 1: Pre-train \std on weak data,  Step 2: Fit \tch to observations from the true function, and Step 3: Fine-tune \std on labels generated by \tch, taking the confidence into account. Red dotted borders and blue solid borders depict components with trainable and non-trainable parameters, respectively.}}
    \label{fig:model}
    \vspace{-10pt}
\end{figure*}

We propose a setting consisting of two main modules. One is called the \std and is in charge of learning a suitable data representation and performing the main prediction task, the other is the \tch which modulates the learning process by modeling the inaccuracies in the labels. 
We explain our approach in much more detail in Section~\ref{sec:proposed-method}, but at a high level it works as follows (see Figure~\ref{fig:model}): We pretrain the student network on weak data to learn an initial task-dependent data representation which we pass to the teacher along with the strong data. The teacher then learns to predict the strong data, but crucially, \emph{based on the student's learned representation}. This then allows the teacher to generate new labeled data from unlabeled data, and in the process correct the student's mistakes, leading to a better data representation and better final predictor. 

\section{\fwlfull (\fwl)}
\label{sec:proposed-method}

In this section, we describe our proposed \fwl approach for semi-supervised learning when we have access to weak supervision (e.g.\ heuristics or weak annotators). We assume we are given a large set of unlabeled data samples, a heuristic labeling function called the \emph{\wa}, and a small set of high-quality samples labeled by experts, called the \emph{strong dataset}, consisting of tuples of training samples $x_i$ and their true labels $y_i$, i.e. $\mathcal{D}_s=\{(x_i,y_i)\}$. We consider the latter to be observations from the true target function that we are trying to learn. 
We use the \wa to generate labels for the unlabeled samples. Generated labels are noisy due to the limited accuracy of the \wa. This gives us the \emph{weak dataset} consisting of tuples of training samples $x_i$ and their weak labels $\tilde{y}_i$, i.e. $\mathcal{D}_w=\{(x_i, \tilde{y}_i)\}$.  Note that we can generate a large amount of weak training data $\mathcal{D}_w$ at almost no cost using the \wa. In contrast, we have only a limited amount of observations from the true function, i.e. $|\mathcal{D}_s| \ll |\mathcal{D}_w|$. 

Our proposed setup comprises a neural network called the \textbf{\std} and a Bayesian function approximator called the \textbf{\tch}. The training process consists of three phases which we summarize in Algorithm~\ref{alg:main} and Figure~\ref{fig:model}.

\setlength{\textfloatsep}{10pt}
\begin{algorithm}[t!]
\fontsize{9}{11}\selectfont
\caption{\small \fwlfull.}
\begin{algorithmic}[1]
\label{alg:main}

\STATE{Train the \std on samples from the weakly-annotated data $D_w$.}
\STATE{Freeze the representation-learning component $\psi(.)$ of the \std and train \tch on the strong data $D_s={(\psi(x_j),y_j)}$. Apply \tch to unlabeled samples $x_t$ to obtain soft dataset $D_{sw}=\{(x_t, \bar{y}_t)\}$ where $\bar{y}_t=T(x_t)$ is the soft label and for each instance $x_t$, the uncertainty of its label, $\Sigma(x_t)$, is provided by the \tch.} 
\STATE{Train the \std on samples from $D_{sw}$ with SGD and modulate the step-size $\eta_t$ according to the per-sample quality estimated using the \tch (Equation~\ref{eqn:eta2}).}
\end{algorithmic}
\end{algorithm}

\textbf{Step 1} \emph{Pre-train the \std on $\mathcal{D}_w$ using weak labels generated by the \wa}. 

The main goal of this step is to learn a \emph{task dependent} representation of the data as well as pretraining the \std. The \std function is a neural network consisting of two parts. The first part $\psi(.)$ learns the data representation and the second part $\phi(.)$ performs the prediction task (e.g. classification). Therefore the overall function is $\hat{y}=\phi(\psi(x_i))$. The \std is trained on all samples of the weak dataset $\mathcal{D}_w=\{(x_i, \tilde{y}_i)\}$. For brevity, in the following, we will refer to both data sample $x_i$ and its representation $\psi(x_i)$ by $x_i$ when it is obvious from the context. 
From the self-supervised feature learning point of view, we can say that representation learning in this step is solving a surrogate task of approximating the expert knowledge, for which a noisy supervision signal is provided by the \wa.

\textbf{Step 2} \emph{Train the \tch on the strong data $(\psi(x_j),y_j) \in \mathcal{D}_s$ represented in terms of the student representation $\psi(.)$ and then use the \tch to generate a soft dataset $\mathcal{D}_{sw}$ consisting of $\langle \textrm{sample} $ $, \textrm{predicted label}, \textrm{ confidence} \rangle$ for \textbf{all} data samples.} 

We use a Gaussian process as the \tch to capture the label uncertainty in terms of the \std representation, estimated w.r.t\ the strong data. A prior mean and co-variance function is chosen for $\mathcal{GP}$. The learned embedding function $\psi(\cdot)$ in Step 1 is then used to map the data samples to dense vectors as input to the $\mathcal{GP}$. 
We use the learned representation by the \std in the previous step to compensate lack of data in $\mathcal{D}_s$ and the \tch can enjoy the learned knowledge from the large quantity of the weakly annotated data. This way, we also let the \tch  see the data through the lens of the \std.

The $\mathcal{GP}$ is trained on the samples from $\mathcal{D}_s$ to learn the posterior mean ${m}_{\rm post}$ (used to generate soft labels) and posterior co-variance $K_{\rm post}(.,.)$ (which represents label uncertainty).
%
We then create the \emph{soft dataset} $\mathcal{D}_{sw}=\{(x_t,\bar{y}_t)\}$ using the posterior $\mathcal{GP}$, input samples $x_t$ from $\mathcal{D}_w \cup \mathcal{D}_s$, and predicted labels $\bar{y}_t$ with their associated uncertainties as computed by $T(x_t)$ and $\Sigma(x_t)$:
\begin{eqnarray*}
\tfunc(x_t) &=& g({m}_{\rm post}(x_t))\\
\Sigma(x_t) &=& h(K_{\rm post}(x_t,x_t))
\end{eqnarray*}
The generated labels are called \emph{soft labels}. Therefore, we refer to $\mathcal{D}_{sw}$ as a soft dataset. $g(.)$ transforms the output of $\mathcal{GP}$ to the suitable output space. For example in classification tasks, $g(.)$ would be the softmax function to produce probabilities that sum up to one. 
For multidimensional-output tasks where a vector of variances is provided by the $\mathcal{GP}$, the vector $K_{\rm post}(x_t,x_t)$ is passed through an aggregating function $h(.)$ to generate a scalar value for the uncertainty of each sample. 
Note that we train $\mathcal{GP}$ only on the strong dataset $\mathcal{D}_s$ but then use it to generate soft labels $\bar{y}_t = \tfunc(x_t)$ and uncertainty $\Sigma(x_t)$ for samples belonging to $\mathcal{D}_{sw}=\mathcal{D}_w\cup \mathcal{D}_s$.

In practice, we furthermore divide the space of data into several regions and assign each region a separate $\mathcal{GP}$ trained on samples from that region. This leads to a better exploration of the data space and makes use of the inherent structure of data. The algorithm called clustered $\mathcal{GP}$ gave better results compared to a single GP.

\textbf{Step 3} \emph{Fine-tune the weights of the \std network on the soft dataset, while modulating the magnitude of each parameter update by the corresponding \tch-confidence in its label.}

The \std network of Step 1 is fine-tuned using samples from the soft dataset $\mathcal{D}_{sw}=\{(x_t, \bar{y}_t)\}$ where $\bar{y}_t=\tfunc(x_t)$.
The corresponding uncertainty $\Sigma(x_t)$ of each sample is mapped to a confidence value according to Equation~\ref{eqn:eta2} below, and this is then used to determine the step size for each iteration of the stochastic gradient descent (SGD). So, intuitively, for data points where we have true labels, the uncertainty of the \tch is almost zero, which means we have high confidence and a large step-size for updating the parameters. However, for data points where the \tch is not confident, we down-weight the training steps of the \std. This means that at these points, we keep the \std function as it was trained on the weak data in Step 1.

More specifically, we update the parameters of the \std by training on $\mathcal{D}_{sw}$ using SGD:
\begin{eqnarray*}
  \pmb{w}^* &=& \argmin_{\pmb{w} \in \mathcal{W}} \> \frac{1}{N}\sum_{(x_t,\bar{y}_t) \in \mathcal{D}_{sw}}l(\pmb{w}, x_t, \bar{y}_t) + \mathcal{R}(\pmb{w}), \\
  \pmb{w}_{t+1} &=& \pmb{w}_t - \eta_t(\nabla l(\pmb{w},x_t,\bar{y}_t) + \nabla \mathcal{R}(\pmb{w}))
\end{eqnarray*}
where $l(\cdot)$ is the per-example loss, $\eta_t$ is the total learning rate, $N$ is the size of the soft dataset $\mathcal{D}_{sw}$, $\pmb{w}$ is the parameters of the \std network, and $\mathcal{R(.)}$ is the regularization term.

We define the total learning rate as $\eta_t = \eta_1(t)\eta_2(x_t)$, where $\eta_1(t)$ is the usual learning rate of our chosen optimization algorithm that anneals over training iterations, and $\eta_2(x_t)$ is a function of the label uncertainty $\Sigma(x_t)$ that is computed by the \tch for each data point. Multiplying these two terms gives us the total learning rate. In other words, $\eta_2$ represents the \emph{fidelity} (quality) of the current sample, and is used to multiplicatively modulate $\eta_1$. Note that the first term does not necessarily depend on each data point, whereas the second term does. We propose
\begin{equation}
 \label{eqn:eta2}
 \eta_2(x_t) = \exp[-\beta \Sigma(x_t)],    
\end{equation}
to exponentially decrease the learning rate for data point $x_t$ if its corresponding soft label $\bar{y}_t$ is unreliable (far from a true sample). In Equation~\ref{eqn:eta2}, $\beta$ is a positive scalar hyper-parameter. Intuitively, small $\beta$ results in a \std which listens more carefully to the \tch and copies its knowledge, while a large $\beta$ makes the \std pay less attention to the \tch, staying with its initial weak knowledge. 
More concretely speaking, as $\beta \to 0$ \std places more trust in the labels $\bar{y}_t$ estimated by the \tch and the \std copies the knowledge of the \tch. On the other hand, as $\beta \to \infty$, \std puts less weight on the extrapolation ability of $\mathcal{GP}$ and the parameters of the \std are not affected by the correcting information from the \tch. 

\section{Experiments}
\label{sec:experiments}

In this section, we apply \fwl to \emph{document ranking}.
We evaluate the performance of our method compared to the following baselines:
\begin{enumerate}[leftmargin=*]
\setlength{\topsep}{0.1pt}
\setlength{\partopsep}{0.1pt}
\setlength{\itemsep}{0.1pt}
\setlength{\parskip}{0.1pt}
\setlength{\parsep}{0.1pt}
\item 
\textbf{WA}. The \wa, i.e. the unsupervised method used for annotating the unlabeled data.
\item
\textbf{NN$_{\text{W}}$}. The \std trained only on weak data.
\item
\textbf{NN$_{\text{S}}$}. The \std trained only on strong data.
\item
\textbf{NN$_{\text{S}^+\text{/W}}$}. 
The \std trained on samples that are alternately drawn from $\mathcal{D}_w$ without replacement, and $\mathcal{D}_s$ with replacement. Since $|\mathcal{D}_s| \ll |\mathcal{D}_w|$, it oversamples the strong data.
\item
\textbf{NN$_{\text{W} \to \text{S}}$}. The \std trained on weak dataset $\mathcal{D}_w$ and fine-tuned on strong dataset $\mathcal{D}_s$.
\item
\textbf{\fwl}. Our \fwl model, i.e.\ the \std trained on the weakly labeled data and fine-tuned on examples labeled by the \tch using the confidence scores.

\end{enumerate}

\subsection{Document Ranking}

This task is the core information retrieval problem and is challenging as the ranking model needs to learn a representation for long documents and capture the notion of relevance between queries and documents. Furthermore, the size of publicly available datasets with query-document relevance judgments is unfortunately quite small ($\sim 250$ queries).
We employ a pairwise neural ranker architecture as the \std~\citep{Dehghani:2017:SIGIR}. 
In this model, ranking is cast as a regression task. Given each training sample $x$ as a triple of query $q$, and two documents $d^+$ and $d^-$, the goal is to learn a function $\mathcal{F} : \{<q, d^+, d^->\} \rightarrow \mathbb{R}$, which maps each data sample $x$ to a scalar output value $y$ indicating the probability of $d^+$ being ranked higher than $d^-$ with respect to $q$.

\begin{figure}{t}
    \centering
            \includegraphics[width=0.45\columnwidth]{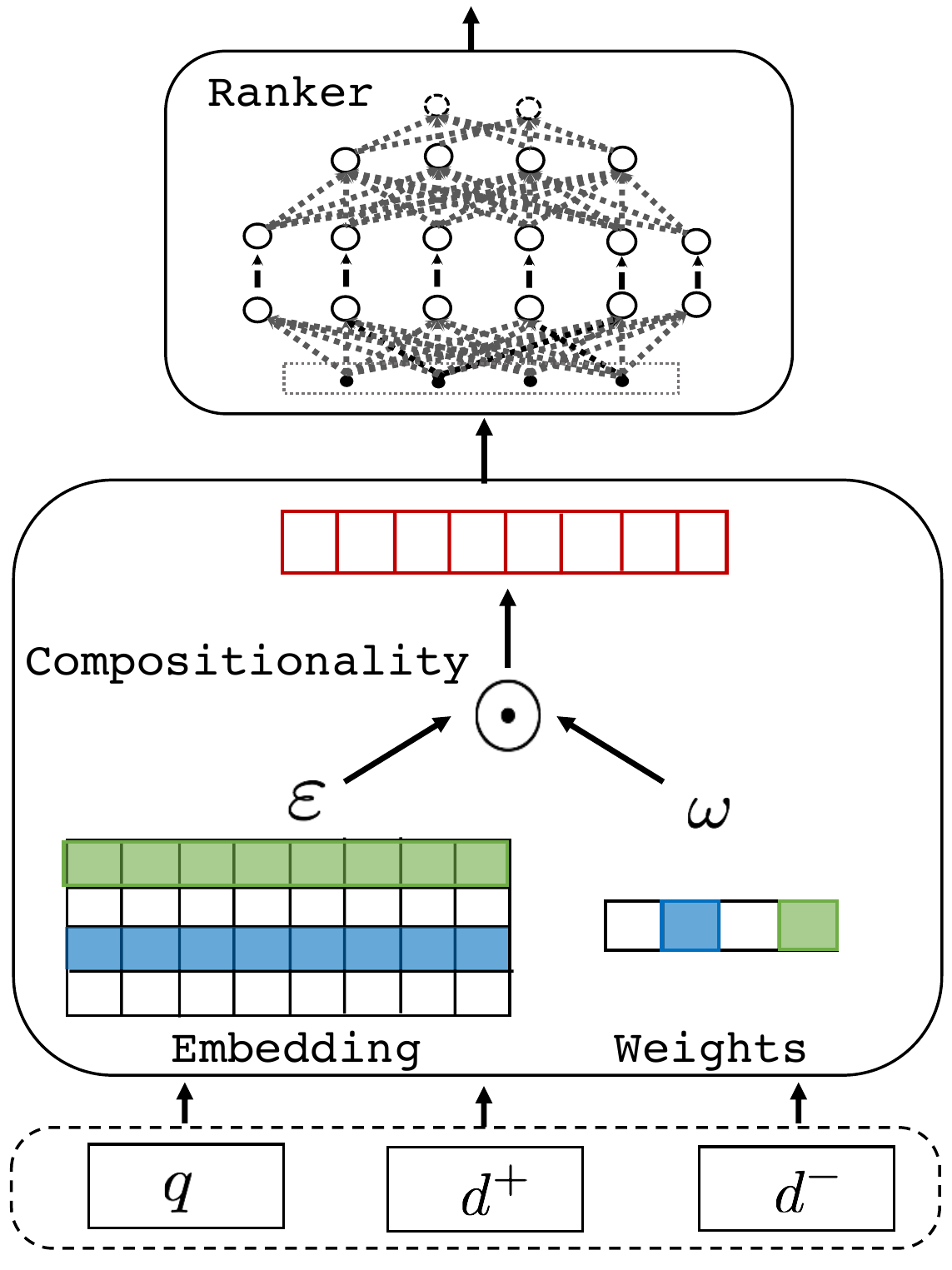}
    \caption{\fontsize{8}{7}\selectfont{The \std for the document ranking task.}}
    \label{fig:ranker}
    \vspace{-5pt}
\end{figure}

\mypar{The \std} follows the architecture proposed in~\citep{Dehghani:2017:SIGIR}. The first layer of the network, i.e. representation learning layer $\psi: \{<q, d^+, d^->\} \rightarrow \mathbb{R}^{m}$ maps each input sample to an $m$-\:dimensional real-valued vector. In general, besides learning embeddings for words, function $\psi$ learns to compose word embedding based on their global importance in order to generate query/document embeddings.
The representation layer is followed by a simple fully-connected feed-forward network with a sigmoidal output unit to predict the probability of ranking $d^+$ higher than $d^-$.
The general schema of the \std is illustrated in Figure~\ref{fig:ranker}.

\mypar{The \tch} is implemented by clustered $\mathcal{GP}$ algorithm.

\mypar{The \wa} is BM25~\citep{Robertson:2009}, a well-known unsupervised method for scoring query-document pairs based on statistics of the matched terms.

\newcommand{\psindex}[1]{\ensuremath{\operatorname{^{\blacktriangleup {#1}}}}}

\begin{table}[tbp]

\scriptsize

\caption{\label{tbl_main} \fontsize{8}{7}\selectfont{Performance of \fwl approach and baseline methods for ranking task. \psindex{i} indicates that the improvements with respect to the baseline $i$ are statistically significant at the 0.05 level using the paired two-tailed t-test with Bonferroni correction.}}
\vspace{-5pt}
\centering
\begin{adjustbox}{max width=\columnwidth}
\begin{tabular}{r l l l l l}
\toprule
& \multirow{2}{*}{Method} &
\multicolumn{2}{c}{Robust04} & \multicolumn{2}{c}{ClueWeb09-CatB\protect\footnotemark}
\\ 
\cmidrule(lr){3-4} \cmidrule(lr){5-6}
& & \small{MAP} & \small{nDCG@20}
& \small{MAP} & \small{nDCG@20}
\\ \midrule
1 & \small{WA$_\text{BM25}$} 
& 0.2503\psindex{3} & 0.4102\psindex{3}  
& 0.1021\psindex{3} & 0.2070\psindex{3}
\\ \midrule
2 & \small{NN$_{\text{W}}$ \citep{Dehghani:2017:SIGIR}} 
& 0.2702\psindex{13} & 0.4290\psindex{13}  
& 0.1297\psindex{13} & 0.2201\psindex{13}
\\
3 & \small{NN$_{\text{S}}$} 
& 0.1790\ & 0.3519\  
& 0.0782\ & 0.1730\
\\ \midrule
4 & \small{NN$_{\text{S}^+\text{/W}}$} 
&  0.2763\psindex{123} & 0.4330\psindex{123} 
&  0.1354\psindex{123} & 0.2319\psindex{123}
\\
5 & \small{NN$_{\text{W} \to \text{S}}$} 
&  0.2810\psindex{123} & 0.4372\psindex{123} 
&  0.1346\psindex{123} & 0.2317\psindex{123}
\\
6 & \small{\fwl}
& \textbf{0.3124}\psindex{12345}  & \textbf{0.4607}\psindex{12345}  
& \textbf{0.1472}\psindex{12345}  & \textbf{0.2453}\psindex{12345} 
\\\bottomrule
\end{tabular}
\end{adjustbox}
\vspace{-5pt}
\end{table}

\mypar{Results and Discussions} We conducted 3-fold cross validation on $\mathcal{D}_s$ (the strong data) with 80/20 training/validation split, and report two standard evaluation metrics for ranking: mean average precision (MAP) of the top-ranked $1,000$
documents and normalized discounted cumulative gain calculated for the top $20$ retrieved documents (nDCG@20). In all of the experiments, the experimental setup and preprocessing are similar to~\citep{Dehghani:2017:SIGIR,Dehghani2017:CIKM}.
Table~\ref{tbl_main} shows the performance on both datasets. As can be seen, \fwl provides a significant boost on the performance over all datasets.
In the ranking task, the \std is designed in particular to be trained on weak annotations~\citep{Dehghani:2017:SIGIR}, hence training the network only on weak supervision, i.e. NN$_\text{W}$ performs better than NN$_\text{S}$. This can be due to the fact that ranking is a complex task requiring many training samples, while relatively few data with true labels are available.

Alternating between strong and weak data during training, i.e. NN$_{\text{S}^+\text{/W}}$ seems to bring little (but statistically significant) improvement. However, we can gain better results by the typical fine-tuning strategy, NN$_{\text{W} \to \text{S}}$. 

\footnotetext{Spam documents were filtered out using the Waterloo spam scorer (\url{http://plg.uwaterloo.ca/~gvcormac/clueweb09spam/})~\citep{Cormack:2011} with the default threshold $70\%$.}

\subsection{Sensitivity to Weak Annotation Quality}
Our proposed setup in \fwl requires defining a so-called ``\wa'' to provide a source of weak supervision for unlabelled data.
Now, in this section, we study how the quality of the weak annotator may affect the performance of the \fwl, for the task of document ranking.

To do so, besides BM25~\citep{Robertson:2009}, we use three other weak annotators: 
\begin{figure}[t]
    \centering
    \includegraphics[width=0.9\columnwidth]{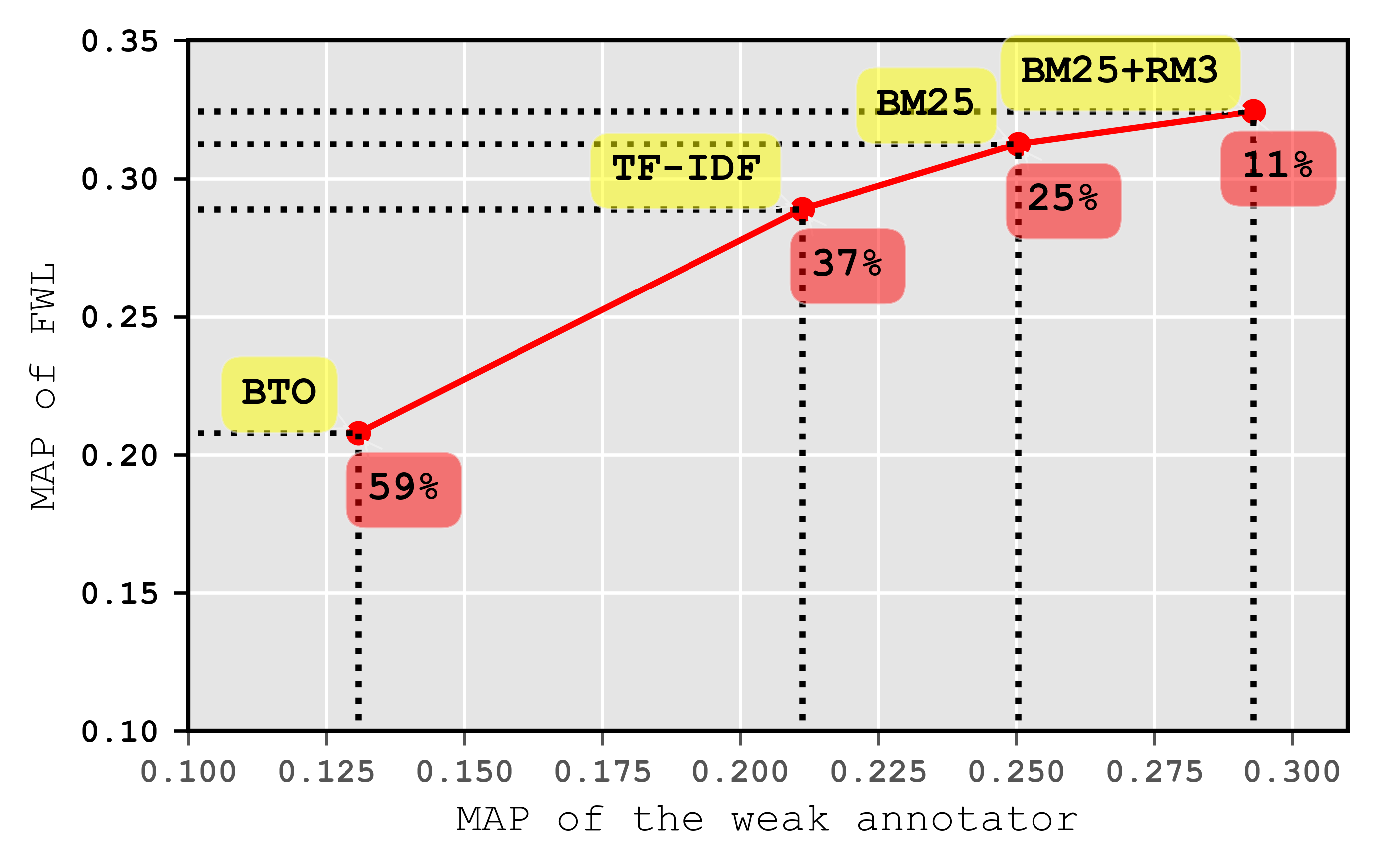}
    \vspace{-10pt}
    \caption{\fontsize{8}{7}\selectfont{Performance of \fwl versus the corespondence \wa in the document ranking task, on Robust04 dataset}.}
    \label{fig:sensitivity}
    \vspace{-1pt}
\end{figure}
vector space model~\citep{salton1973specification} with binary term occurrence (BTO) weighting schema and vector
space model with TF-IDF weighting schema, which are both weaker than BM25, 
and BM25+RM3~\citep{Abdul-jaleel:2004} that uses pseudo-relevance feedback~\citep{Dehghani:2016:cikm-long} on top of BM25, leading to better labels. 

Figure~\ref{fig:sensitivity} illustrates the performance of these four \was in terms of their mean average precision (MAP) on the test data, versus the performance of \fwl given the corresponding \wa. As it is expected, the performance of \fwl depends on the quality of the employed \wa.
The percentage of improvement of \fwl over its corresponding \wa on the test data is also presented in Figure~\ref{fig:sensitivity}. As can be seen, the better the performance of the \wa is, the less the improvement of the \fwl would be.

\section{Conclusion}
\label{sec:conclusion}

Training neural networks using large amounts of weakly annotated data is an attractive approach in scenarios where an adequate amount of data with true labels is not available, a situation which often arises in practice.
In this paper, we make use of \fwlfulllc (\fwl), a new student-teacher framework for semi-supervised learning in the presence of weakly labeled data.
We applied \fwl to document ranking and empirically verified that \fwl speeds up the training process and improves over state-of-the-art semi-supervised alternatives.

Our general conclusion is that explicitly modeling label quality is both possible and useful for learning task dependent data representations.  The student-teacher configuration conceptually allows us to distinguish between the role of the student who learns the target representation, and the teacher who both learns to estimate the confidence in labels as well as adapts to the needs of the student by taking the student's current state into account.  One key observation is that the model with explicit feedback about the strong labels (i.e., pre-training and fine-tuning) is not as effective as the teacher model that implicitly gives feedback from the strong labels in terms of label confidence.  While this can be explained in terms of avoiding overfitting or loss of generalization, there is also a conceptual explanation in terms of promoting the student to learn by not overruling or correcting, but by giving the student the right feedback to allow for a learning by discovery approach, retaining the full generative power of the student model.  Arguably, this even resembles a kind of ``Socratic dialog'' between student and teacher.

\smallskip
\mypar{Acknowledgments}  \small This research is funded in part by the Netherlands Organization for Scientific Research (NWO;  ExPoSe project, NWO CI \# 314.99.108). 

\vspace*{-\parskip}

\bibliographystyle{abbrvnat} 
\bibliography{sigproc} 
\end{document}